# Pressure Ulcers and Dressings: A Strain Sensitivity Analysis of the Boundary Conditions of a Finite Element Model

**Nolwenn Fougeron** [1], **Isabelle Rivals** [2,3], **Nathanaël Connesson** [1], **Grégory Chagnon** [1], **Thierry Alonso** [1], **Laurent Pasquinet** [4], **Stéphane Auguste** [4], **Antoine Perrier** [1,5] and **Yohan Payan** [1]

[1] Laboratoire TIMC-IMAG, Univ. Grenoble Alpes, CNRS, UMR 5525, VetAgro Sup, Grenoble INP, TIMC, 38000 Grenoble, France
[2] UMR8256 Biological Adaptation and Ageing Research Group, Sorbonne Université, INSERM, UMRS1158, Neurophysiologie Respiratoire Expérimentale et Clinique, 75013 Paris, France
[3] Équipe de Statistique Appliquée, ESPCI Paris, PSL Research University, UMRS1158, 75005 Paris, France
[4] Urgo Research, Innovation & Development, 21300 Chenôve, France
[5] Département de Médecine de L'adolescent, Sorbonne Université Médecine, Assistance Publique Hôpitaux de Paris (APHP), Service de Diabétologie, Hôpital Pitié-Salpêtrière, 75013 Paris, France

**Abstract:** Recently, a new bi-layer dressing was proposed by Urgo RID to reduce the healing time of pressure ulcers (PU). This dressing was numerically evaluated in previously published work. In the current work, the influence on the maximal shear strains of modelling parameters such as the dressing local geometry, the pressure applied by the gauze inside the wound, the wound deepness, and the mattress stiffness, was assessed. A sensitivity analysis was performed on these four parameters. Among all experiments, the mean maximal Green–Lagrange shear strain was 0.29. The gauze pressure explained 60% of the model response in terms of the volume of tissues under strains of 0.3, while the wound deepness explained 28%. The mattress had a significant, but low impact, whereas the dressing local geometry had no significant impact. As expected, the wound deepness was one of the most influential parameters. The gauze turned out to be more significant than expected. This may be explained by the large range of values chosen for this study. The results should be extended to more subjects, but still suggest that the gauze is a parameter that might not be neglected. Care should also be taken in clinical practice when using gauze that could have either a positive or negative impact on the soft tissues' strains. This may also depend on the wound deepness.

## 1. Introduction

Pressure ulcers (PU) are injuries to the skin and underlying tissues that are common adverse events in healthcare. For example, in intensive care units, PU prevalence reaches almost 27% [1]. In any healthcare facility, the risk of developing a PU is increased for older patients, patients with spinal cord injuries, or comorbidities [2]. PU have terrible consequences on the quality of life of patients including longer hospitalisation time, social isolation, and pain [3,4].

PU are localised wounds that propagate in the soft tissues after a detrimental external loading. They are classified from stage-1, for light wounds, to stage-4, for the most severe wounds. Short time (some minutes), but intense, load application is sufficient to cause a PU, while reduced loads applied for an extended period (2 to 4 h) can also lead to this kind of wound [5]. They have a multifactorial origin, but mechanical loads applied to the tissues are considered to play a significant role in the onset of PU. Pressure or shear loads applied at the skin level may lead to significant internal strains [6]. Strains and, more particularly, the Green–Lagrange maximal shear strains, appeared to be a mechanical biomarker for the development of PU [7]. When these strains exceed the cell's ability to deform, in most cases under bony prominences, this eventually leads to cell death and the development of a wound [7–9]. The sacrum is the most affected area of the human body. In this case, the recommended procedure to treat PU consists of the unloading of the weakened tissues, which can be tedious to do continuously, particularly if several PU are present. Dressings are common medical devices used to improve the healing process of PU and a huge range of products are proposed to clinicians. Yet, the mean healing time of PU is estimated to be 3 weeks and can sometimes exceed 10 weeks [10]. Recently, Urgo RID developed a new concept of dressing to improve the healing of PU by reducing internal strains. This dressing consists of two layers, the first one being the classic Urgo Start Plus Border dressing and the second one consisting of an unloading material. This material is cut into alveoli that can be removed under the wound to relocate the loads outside of the wound region when complete unloading of the PU is not temporally possible. The ability of this dressing to alleviate soft tissues has already been studied previously [11]; however, question marks remain about the use of the dressing. The impact of the dressing with different wound deepness, quantities of alveoli removed, or mattress stiffnesses still needs to be estimated. Furthermore, the interaction with the gauze that is sometimes applied in the wound to absorb part of the exudate has not been studied yet.

Finite element modelling is a common method applied to compute the soft tissues' internal strains. Ceelen et al. [12] proposed and validated this method on rat models for the estimation of the Green–Lagrange maximal shear strains. Yet, few models of the sacrum region were proposed in the literature [13]. These models were mainly proposed to compute internal and external stresses in soft tissues without PU. Some studies also applied the finite element modelling method to evaluate penetrating ulcers in the cardiovascular domain [14], yet few efforts were made for PU in soft tissues such as skin or adipose tissues. To the authors' knowledge, the group of Amit Gefen (Tel Aviv University) was the only one to propose a finite element model of the injured tissues with a stage-4 PU. The authors showed that adding a multilayer dressing allowed the reduction of internal and external stresses around the wound. Several other studies from that group also performed comparative analyses of the finite element model of the sacrum region with various dressings or mattresses. They compared the use of silicone foam dressing with various material parameters and showed that dressings anisotropy helped reduce the internal and external stresses [15]. In another study, the authors from this group showed that the increase in mattress stiffness induced an increase in internal stresses [16]. They also studied various soft cellulose fluff core dressings with two mattress conditions and two moisture states of the dressings. Few differences could be noted among the dressings, but better performances were obtained with the softest mattress [17,18]. These studies bring interesting insights into how the change of boundary conditions may impact the response of the soft tissue and potentially the onset or propagation of a PU. However, none of the previous studies reported statistical analysis on the relative importance of the studied parameters [19]. Furthermore, the gauze inside the wound has still not been modelled.

The current study aims to estimate the relative impact of the dressing geometry, mattress stiffness, use of gauze, and PU deepness on the soft tissues' maximal shear strains around the wound. A sensitivity analysis was performed on these four parameters using a previously designed parametric model of the sacrum region.

## 2. Materials and Methods

*2.1. Reference Finite Element Model*

To reduce the computation time, a parametric approach was adopted. The model consisted of several layers: the skin, adipose tissues, both dressing layers, and a mattress. The skin and adipose tissue thicknesses were set to 1.30 mm and 22.30 mm, respectively [19,20]. To simulate a bony prominence on the median crest of the sacrum, an imprint of the sacrum geometry was approximated by a portion of a sphere with an ellipsoidal volume on top of it. The adipose tissue thickness was thus reduced to 13.30 mm under the bony prominence. The sacrum bone was set as rigid with the pilot node at the centre of the area. A PU from stage-2 to stage-3 was added to the model, with various depths defined after, while the radius was set to 15.00 mm. The dressing was modelled with two layers referred to as dressing layer 1 and dressing layer 2 (Figure 1). Dressing layer 1 is the unloading material cut into alveoli that is in contact with the mattress, and dressing layer 2 is the UrgoStart Plus Border dressing that is in contact with the skin. Both layers were modelled as a cylindrical layer with a radius of 125.00 mm. The thickness of dressing layer 2 was set to 3.50 mm, whereas the thickness of dressing layer 1 was set to 5.20 mm. A mattress with a height of 50.00 mm was added to the model. The diameter of the model was 250.00 mm to avoid boundary effects in the wound area. Symmetry in the sagittal plane was also considered so only half of the model was used for the simulation (Figure 2). All components were meshed with SOLID185 linear hexahedral elements ANSYS APDL (ANSYS 2020 R2 software, ANSYS Inc., Cannonsburg, PA, USA) with a mixed pressure-displacement formulation for the soft tissues. The model was composed, at most, of 6088 elements.

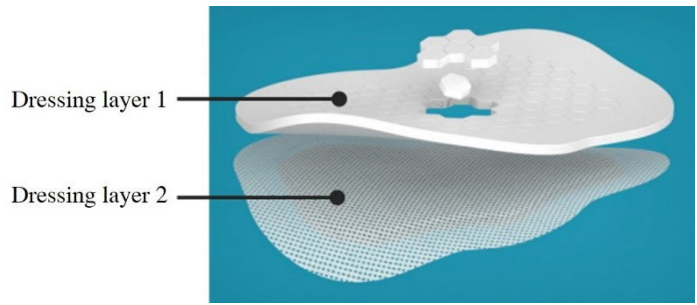

**Figure 1.** The new dressing design developed by Urgo RID.

The dressing layers were tied together. Tie constraints were also used between the soft tissue layers and between the skin and dressing layer 2. A coefficient of friction of 0.62 was defined between dressing layer 2 and the mattress. This value was computed from friction tests performed at Urgo RID. The dressing, glued on a calibrated weight, was positioned on a rigid plate cover with a clinical sheet. A gradually increasing force was applied to a cable attached to the dressing. The coefficient of friction was the ratio of the force that pulled the dressing and the calibrated weight. Between the skin and the mattress, this coefficient was set to 0.43 [20]. A vertical force of 217 N was applied to the pilot node of the sacrum area, as illustrated in Figure 2a. Considering the symmetry of the model, this corresponded to 47% of the bodyweight of a 94 kg subject [21]. The bottom nodes of the mattress were fixed in position. Simulations were performed with ANSYS in a quasi-static analysis with an implicit scheme.

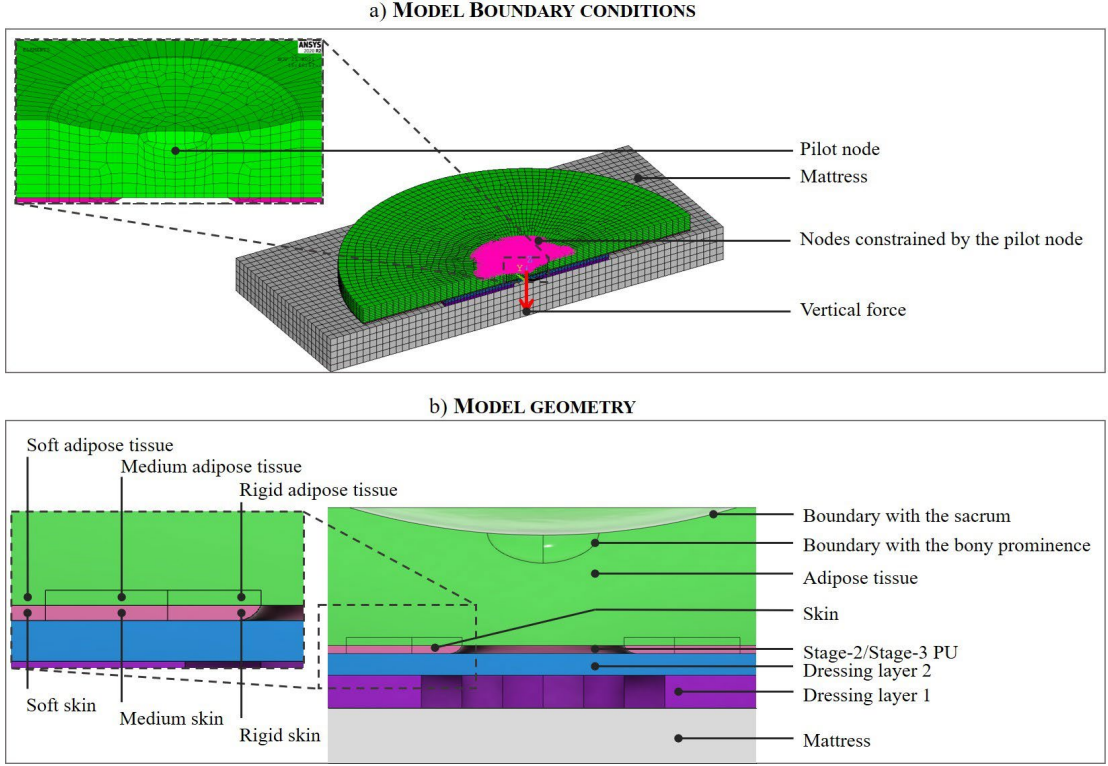

**Figure 2.** Model geometry and boundary conditions.

Soft tissues were modelled with non-linear hyperelastic isotropic constitutive equations. More particularly, the skin was modelled with the law proposed by Isihara et al. [22]. The material parameters were optimised using a curve-fitting method from the data of Ni Annaidh et al. [23]. The adipose tissues were modelled with the equation developed by Yeoh [24] with parameters fitted according to the data of Sommer et al. [25]. The soft tissue stiffness was increased close to the wound region, as detailed in Fougeron et al. [11], to account for the stiffening of the tissues surrounding a PU. The constitutive equation for the different tissues is:

$$W = \sum_{i=1}^{n} C_{i0}(\overline{I_1} - 3)^i + \sum_{k=1}^{n} \frac{1}{d_k}(J - 1)^{2k}$$

$$d_1 = d_2 = d_3 = \frac{3(1 - 2\nu)}{2C_{10}(1 + \nu)}$$

where $W$ is the strain energy density function, $C_{i0}$ the material parameters, $\overline{I_1}$ the first deviatoric invariant of the right Cauchy–Green deformation tensor, $J$ the Jacobian of the deformation gradient, and $d_k$ the nearly incompressibility parameters expressed with the Poisson's ratio $\nu$ by the formula provided by Mott et al. [26]. The indices $i$ and $k$ are between 1 and 3 for the skin and between 1 and 2 for the adipose tissues. Soft tissue stiffening was considered by multiplying the $C_{10}$ parameters of the skin and the adipose tissue by a coefficient of 1.0, 1.5, and 2.0 for the soft, medium, and stiff areas, respectively, as detailed in Figure 2.

The value of Poisson's ratio was set to 0.4999 to account for the nearly incompressibility of the soft tissues. Dressing layer 2 was modelled with a linear elastic orthotropic material, whereas layer 1 was defined as a compressible material and modelled with a Blatz–Ko constitutive equation [27].

$$W = \frac{\mu}{2}\left(\frac{I_2}{I_3} + \sqrt{I_3} - 5\right)$$

where $I_2$ and $I_3$ are the second and third invariants of the right Cauchy–Green deformation tensor, respectively, and $\mu$ is the initial shear modulus.

The initial shear modulus µ of dressing layer 1 and Young's moduli of dressing layer 2 were computed from compression and tension tests. According to the literature data, the Poisson ratio of dressing layer 2 was set to 0.2560. The mattress was modelled as a linear elastic isotropic material with a Poisson ratio of 0.3000 and a reference Young modulus, $E$, of 0.23 MPa. The material parameters are detailed in Table 1. Further details about the reference model are provided in Fougeron et al. [11].

Table 1. Material parameters of the model's components.

| Component | $C_{10}$ (MPa) | $C_{20}$ (MPa) | $C_{30}$ (MPa) | $\mu$ (MPa) | $E_X$ (MPa) | $E_Y$ (MPa) | $E_Z$ (MPa) | $d_1$ (MPa$^{-1}$) | $\nu$ |
|---|---|---|---|---|---|---|---|---|---|
| Adipose tissue | $1.3 \times 10^{-4}$ | 0.0 | $12.2 \times 10^{-3}$ | - | - | - | - | 1.6 | 0.4999 |
| Skin | $2.7 \times 10^{-1}$ | 1.9 | - | - | - | - | - | - | 0.4999 |
| Dressing layer 1 | - | - | - | $1.0 \times 10^{-3}$ | - | - | - | - | - |
| Dressing layer 2 | - | - | - | - | 4.4 | 1.8 | $2.6 \times 10^{-2}$ | - | 0.2560 |
| Mattress | - | - | - | - | $2.3 \times 10^{-1}$ | - | - | - | 0.3000 |

*2.2. Sensitivity Analysis*

Principal stretches $\lambda_1$, $\lambda_2$, and $\lambda_3$ were extracted to compute the Green–Lagrange principal strains (Equation (4)). The maximal shear strain, $E_{shear}$, was calculated as detailed in Equation (5).

$$E_i = \frac{1}{2}(\lambda_i{}^2 - 1), i \in [1, 2, 3]$$

$$E_{shear} = \frac{1}{2}\max(|E_1 - E_2|, |E_2 - E_3|, |E_3 - E_1|)$$

Green–Lagrange maximal shear strains are recognised as potential mechanical biomarkers to study the onset and development of PU [7]. In the current study, a region of interest (ROI) was defined for the computation of the strains. The ROI included soft tissues under the wound and in the perilesional area within three times the radius of the PU. Experiments performed on rats suggested the possibility to define a threshold of damage that should be subject-specific [7]. Due to the lack of data on human subjects, the threshold was arbitrarily fixed to 0.3 considering that $E_{shear}$ was below this threshold for healthy tissues.

A sensitivity analysis was performed to assess the relative significance of the model parameters on the volume of healthy tissues. The finite element model was emulated with a polynomial model detailed after, following the method described by Macron et al. [19], to investigate the impact of the following parameters on the volume of healthy tissues: wound deepness, alveoli cutting size, mattress stiffness, and pressure applied by the gauze. The parameters varied between their minimal and maximal values, as detailed in Table 2. After normalisation in [−1, 1], experimental points were chosen according to a three-level full factorial design resulting in 3⁴ combinations (i.e., 81 simulations). Based on the knowledge of expert clinicians, the wound deepness extrema were set to 1.30 mm and 5.00 mm to respectively account for a stage-2 and a stage-3 PU. The recommendations from Urgo about the use of the bi-layer dressing are to remove alveoli around the wound, so this was used as the mean level of the parameter. Then, a layer of alveoli around the wound was added or removed to respectively define the minimal and maximal levels (Figure 3). The mattress stiffness limits were defined according to literature values [16,28]. In the clinical routine, the pressure applied by the gauze may significantly vary depending on its saturation in fluid and on the person who inserts the gauze in the wound. As a consequence, it was chosen to model the effect of the gauze by the pressure applied on the wound walls rather than that of the gauze itself. A finite element preliminary study was performed to define the gauze pressure values. To this end, the volume of healthy tissues was analysed with a 5.0 mm deep PU model for multiple values of pressure between 0.00 MPa and 0.08 MPa. A local optimum was found at 0.02 MPa. As a consequence, the minimal and maximal values were set to 0.00 MPa (i.e., no gauze in the wound) and 0.04 MPa.

**Table 2.** Parameters' minimal, intermediate, and maximal values used as levels for the experimental points of the sensitivity analysis.

| Parameters | Minimal Level | Intermediate Level | Maximal Level |
|---|---|---|---|
| Wound deepness | 1.30 mm | 3.20 mm | 5.00 mm |
| Alveoli cut | Recommended +1 layer | Recommended | Recommended −1 layer |
| Mattress stiffness | 0.03 MPa | 0.23 MPa | 0.43 MPa |
| Gauze pressure | 0.00 MPa | 0.02 MPa | 0.04 MPa |

Given that the local finite element model is rather a qualitative model, a full polynomial model of degree two was considered sufficient to emulate it:

$$y(\theta) = \theta_0 + \sum_{i=1}^{m} \theta_i x_i + \sum_{i=1}^{m} \theta_{ii} x_i^2 + \sum_{i=1}^{m}\sum_{j>i}^{m} \theta_{ij} x_i x_j$$

where $y$ is the volume of healthy tissues, $m$ the number of parameters, $x_i$ the value of the $i$th parameter, and $\theta$ the vector of the adjustable coefficients, which was estimated with ordinary least squares. The value of two for the degree will be further justified in the results section. The sensitivity of the model to each input (linear term, square, order-two interaction) can be simply defined as the percentage of variance due to this input. Assuming, for simplicity, the $m = 4$ parameters independent and uniformly distributed in [−1, 1] (i.e., with second- and fourth-order moments of respectively 1/3 and 4/45), it becomes:

$$\begin{cases} s_i = \mathrm{var}(\theta_i x_i) = \theta_i^2 \mathrm{var}(x_i) = \theta_i^2 \times \frac{1}{3} \\ s_{ii} = \mathrm{var}(\theta_{ii} x_i^2) = \theta_{ii}^2 \mathrm{var}(x_i^2) = \theta_{ii}^2 \times \frac{4}{45} \\ s_{ij} = \mathrm{var}(\theta_{ij} x_i x_j) = \theta_{ij}^2 \mathrm{var}(x_i)\mathrm{var}(x_j) = \theta_{ij}^2 \times \frac{1}{9} \\ \mathrm{var}(y) = \sum_{i=1}^{m} s_i + \sum_{i=1}^{m} s_{ii} + \sum_{i=1}^{m}\sum_{j>i}^{m} s_{ij} \end{cases}$$

The sensitivities to the $i$th parameter and to its interaction with parameter $j$ are hence given by the percentages:

$$S_i = \frac{s_i + s_{ii}}{\mathrm{var}(y)}, S_{ji} = \frac{s_{ij}}{\mathrm{var}(y)} \tag{8}$$

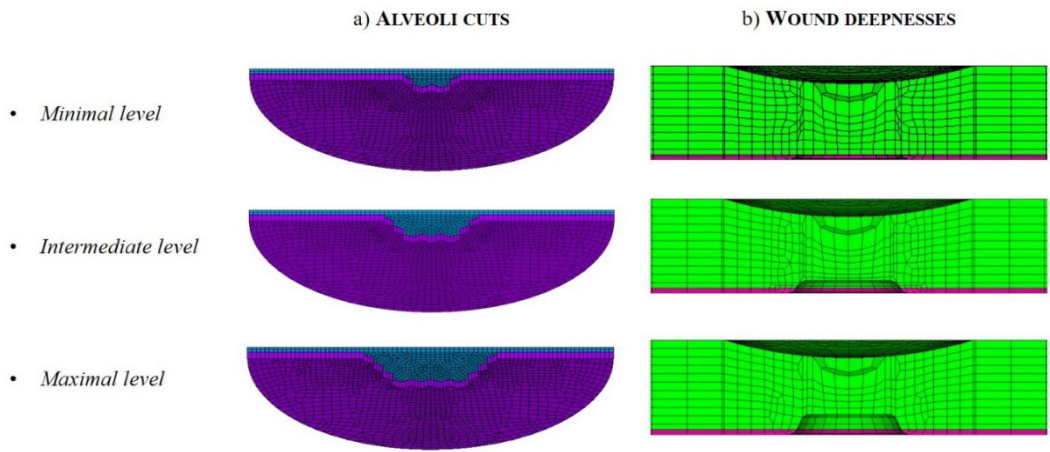

**Figure 3.** Minimal, intermediate, and maximal levels of the alveoli cutting (**a**) and wound deepness (**b**) parameters.

## 3. Results

This section may be divided by subheadings. It should provide a concise and precise 208 description of the experimental results, their interpretation, as well as the experimental 209 conclusions that can be drawn in Table 3.

**Table 3.** Parameter coefficients and polynomial model sensitivities (>1%) in decreasing order of magnitude.

| Parameters | Coefficients $\theta i$ and $\theta ii$ or $\theta ij$ | Sensitivities $S_i$ or $S_{ij}$ (%) |
| --- | --- | --- |
| Gauze pressure | −3.9, −10.7 | 60 |
| Wound deepness | −4.3, −3.3 | 28 |
| Wound deepness*Gauze pressure | +4.6 | 10 |
| Mattress stiffness | +1.1, −0.9 | 1 |

One may notice that approximately 99% of the model response $y$ was explained by four parameters: the gauze pressure, the wound deepness, the interaction of the wound deepness and the gauze pressure, and the mattress stiffness. More particularly, the gauze pressure explained about 60% of the model response, as illustrated in Figure 4a. Considering dressing layer 1, this layer was shown to reduce the maximal shear strains on one model of a stage-2 PU in a previous study. When close enough to the recommended (i.e., plus or minus one layer of alveoli), the change in the volume of healthy tissues was not significant, as presented in Figure 4. On the contrary, wound deepness was a significant parameter that explained 28% of the response (cf. Figure 4). As expected, the interaction of the wound deepness and the gauze pressure was also important, whereas the mattress stiffness had a significant, but low impact on the volume of healthy tissues (cf. Figure 4). Extreme values of gauze pressure seem to have a negative impact on the volume of healthy tissues (cf. Figure 4), suggesting that an optimal value can be found. Tissues around deep PU tend to have more important strains (cf. Figure 4) and softer mattresses may not be suitable in all cases, since the interquartile range of the volume of healthy tissues is larger than for stiffer mattresses. Worst-case scenarios were defined as the 10% experiments with the highest peak maximal shear strains. Among these nine experiments, the peak maximal shear strains were greater than 0.80 and all were designed with the softest mattress and the maximal gauze pressure with various wound deepness and alveoli cut.

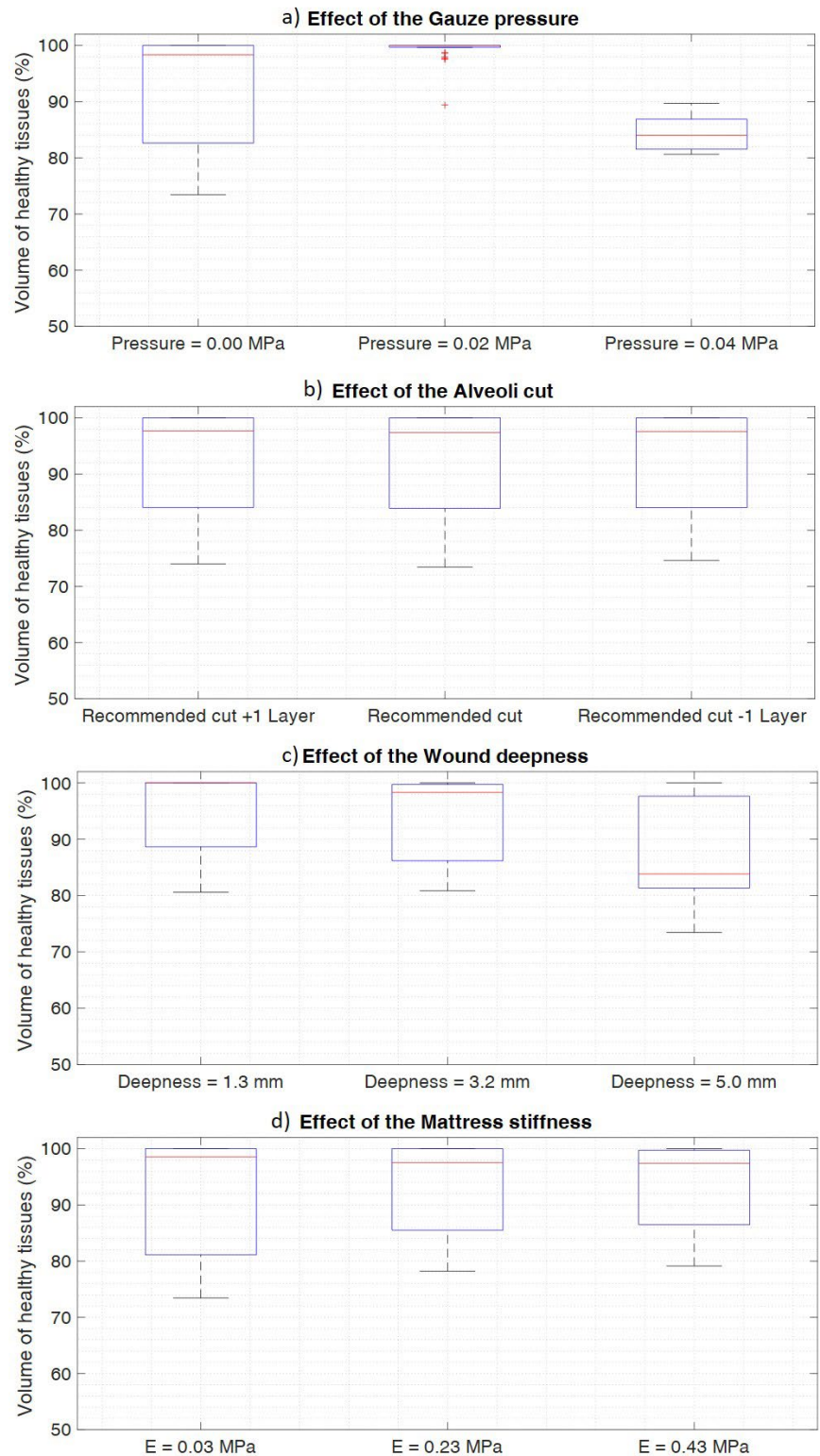

**Figure 4.** Effect of the four parameters on the volume of healthy tissues (i.e., tissues with strains lower than 0.3), with the other three parameters being set to their intermediary value.

To illustrate the results, Green–Lagrange maximal shear strains in the ROI were plotted for some experiments in Figure 5.

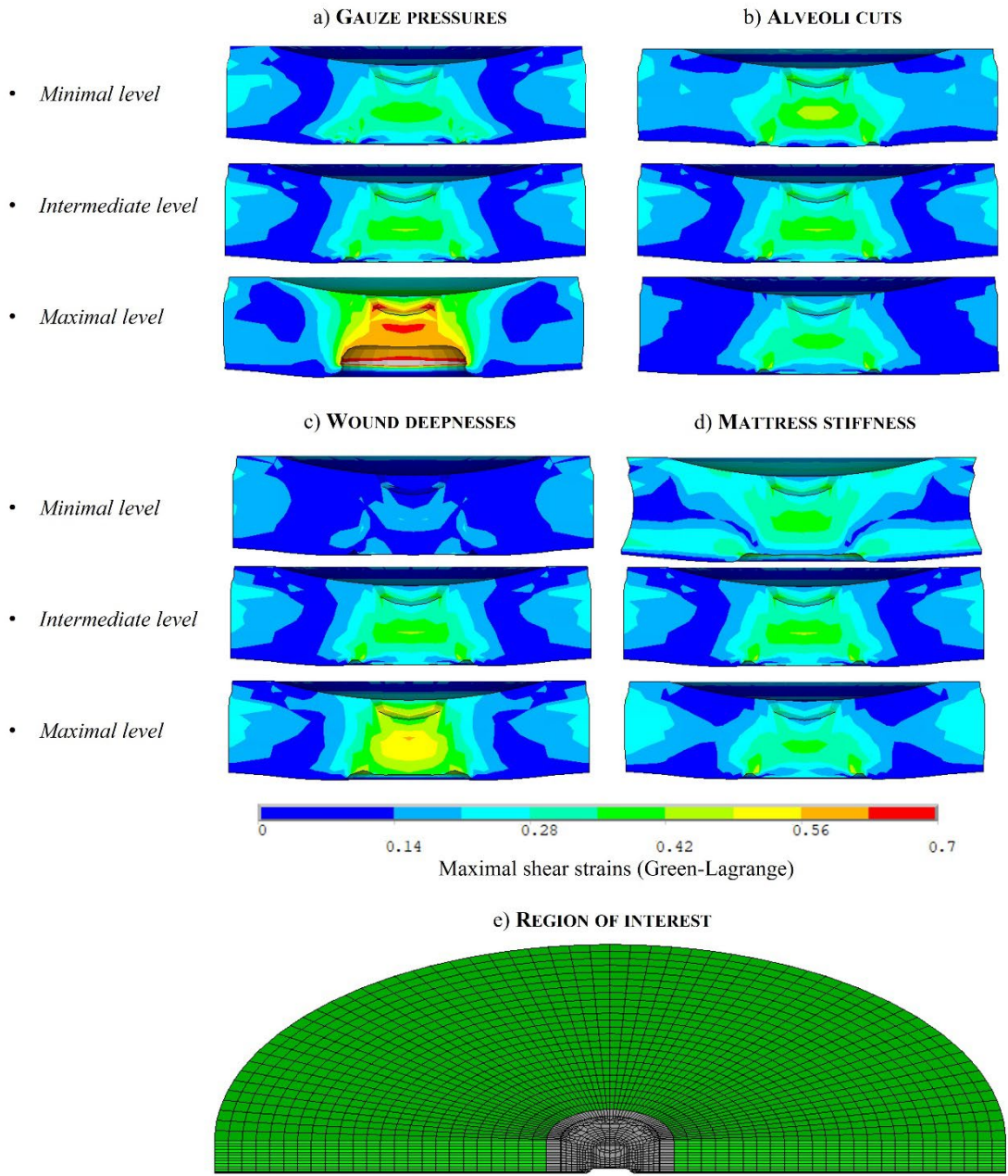

**Figure 5.** Green–Lagrange maximal shear strains in the ROI of some experiments. All parameters were set to the intermediate values except for one that varied according to the defined levels: (**a**) changes in the gauze pressure, (**b**) changes in the alveoli cut, (**c**) changes in the wound deepness, and (**d**) changes in the mattress stiffness. The ROI appears in grey in (**e**).

## 4. Discussion

A new bi-layer dressing has been proposed by Urgo RID to improve the healing of PU. This dressing has previously been studied to evaluate its mechanical impact on the soft tissues in one specific scenario. In this case, the use of the dressing allowed the reduction of internal strains around the wound. Yet, some factors may affect the conclusions: the dressing alveoli cutting, the pressure applied by the gauze inside the wound, the deepness of the wound, and/or the stiffness of the mattress. Thereby, the present study aimed to evaluate the relative importance of these parameters regarding the maximal shear strains around the PU. A sensitivity analysis was performed following a three-level full factorial design.

Among all experiments, the mean maximal shear strain was 0.29 and the peak value reached 0.97. The experiments that reached the highest values of maximal shear strains were all designed with the softest mattress and the maximal gauze pressure. The strain values are in range with the previously published results, but are lower than those obtained by Macron et al., for whom peak values ranged between 1.42 and 4.14. Macron et al. studied the strains under the ischial tuberosities in subjects in a sitting position, which may explain the differences [19]. The computation of the peak maximal shear strain is also local and thus highly sensitive to mesh quality and model non-linearities. Therefore, the volume of healthy tissues was preferred here as a discriminant measure for the sensitivity analysis. The gauze pressure alone explained 60% of the model response, while the wound deepness and the interaction between the gauze pressure and the wound deepness accounted for 28% and 10% of the response, respectively. To the authors' knowledge, this study is the first attempt to assess the impact of these two parameters on the computation of the strains where they both significantly impacted the results. The mattress also had a significant, but low impact. Contrary to the previous studies of Linder-Ganz and Gefen [16], the softest mattress did not necessarily reduce the strains in the ROI. This may be due to the use of the bi-layer dressing in this particular study, which adds a cushion layer between the soft tissues and the mattress. Furthermore, the local approach proposed in this study may not be able to capture the impact of the mattress on a large scale, since weight-bearing areas are limited here. It is worth noting that the results could be affected by the levels chosen for the sensitivity analysis. Mattress stiffness is highly dependent on the brand and few data are provided by the manufacturers. The mattress was modelled with linear elastic homogeneous isotropic material properties, which may not be appropriate for all mattress technologies. The use of gauze was modelled as a homogeneous pressure applied inside the PU. Various products are used by clinicians and the filling of the gauze inside the wound is highly dependent on the operator and the exudate of the wound. The use of pressure allows one to model the effect of the gauze without the need to model all types of commercialised products or operators' protocols. The wound deepness is a significant parameter with an important impact, but in the present study, PU 5.3 mm deep at most were designed. Consequently, the conclusion might not be extrapolated to deeper PU. Other parameters could also have been included in the sensitivity analysis. A geometrical description of the PU such as its diameter or the interaction between the PU diameter and the dressing alveoli cutting could modify the strain distribution. Subject-specific parameters were also not studied in this work. As detailed by Macron et al. [19], materials and thicknesses of soft tissues as well as bone geometries may have a significant impact on strain computation [19]. The material parameters of soft tissues were estimated from cadaveric tests of the literature. Therefore, the current study does not account for the variability of the constitutive behaviours that are proposed in the literature [13,29,30]. The Poisson ratio was also higher than in most literature studies, but this is in range with the recommendation of Bonet and Wood [29] to be close to incompressibility. The soft tissue thicknesses were fixed in the current study even though values from 4.0 mm to 33.5 mm were reported by Clark et al. [30]. Yet, considering all of the parameters would have entailed too many experiments. As a result, it was decided for this study to focus on one particular case for which the model was previously experimentally evaluated, and to evaluate the parameters relating to the use of the dressing in this particular environment: the alveoli cutting, the gauze pressure, the wound deepness, and the mattress stiffness. The present study was not exhaustive on the studied parameters. Further analyses are necessary to include subject-specific parameters obtained on healthy subjects, but also on subjects with PU. The threshold of the strains used to define healthy tissues could also have an impact on the results. Thus, the same sensitivity analysis was performed with a threshold of 0.65 as prescribed by Ceelen et al. [7]. Small discrepancies, a few percent, were noted in terms of sensitivities, but the relative order of the parameters remained the same.

Finally, the results presented here suggest that care should be taken when filling the wound with gauze. Gauze is important to maintain an optimal environment in the wound, particularly in terms of moisture. However, gauze should not be crammed into the wound or filled with too much fluid at the risk of applying too much pressure inside the wound and thus exacerbating the deformations of already weakened soft tissue. Furthermore, as was expected, the deeper the wound, the more strains. Even though the unloading of soft tissues is always prescribed for PU, special care should be taken when dealing with stage-2 and higher PU. To consolidate the conclusion, future work will include the transfer of the proposed modelling on realistic subject-specific geometries of the sacrum and the heel in several patients. This study is a first attempt to numerically evaluate the effect of new dressing designs and to potentially propose guidelines to industrials and clinicians for the use of these medical devices.

**Conflicts of Interest:** This study was financially supported by Urgo RID.